\pdfoutput=1

\documentclass[conference]{IEEEtran}
\usepackage{amsmath,amsfonts,amssymb,mathabx,color}
\usepackage[utf8]{inputenc} 
\usepackage[T1]{fontenc}    
\usepackage{hyperref}       
\usepackage{url}         
\usepackage{booktabs}      
\usepackage{nicefrac}       
\usepackage{microtype}      
\usepackage{color}
\usepackage{bm, graphicx, caption, subcaption}
\usepackage{float}
\usepackage{lipsum}
\usepackage[export]{adjustbox}
\usepackage{bmpsize}

\newcommand{\eqdef}{\coloneq}     
\newcommand{\R}{\mathbb R}
\newcommand{\mrm}{\mathrm}

\ifCLASSINFOpdf
\else
\fi

\begin{document}

%
\title{On the information in spike timing: neural codes derived from polychronous groups}

\author{\IEEEauthorblockN{Zhinus~Marzi, Jo\~ao Hespanha and Upamanyu Madhow}
\IEEEauthorblockA{Department of Electrical and Computer Engineering\\
University of California, Santa Barbara\\
Email: \{zh\_marzi, hespanha, madhow\}@ece.ucsb.edu}
}


%


\maketitle

\begin{abstract}
There is growing evidence regarding the importance of spike timing in neural information processing, with even a small number of spikes carrying information, but computational models lag significantly behind those for rate coding.  Experimental evidence on neuronal behavior is consistent with the dynamical and state dependent behavior provided by recurrent connections.  This motivates the minimalistic abstraction investigated in this paper, aimed at providing insight into information encoding in spike timing via recurrent connections. We employ information-theoretic techniques for a simple reservoir model which encodes input spatiotemporal patterns into a sparse neural code, translating the polychronous groups introduced by Izhikevich into codewords on which we can perform standard vector operations.  We show that the distance properties of the code are similar to those for (optimal) random codes. In particular, the code meets benchmarks associated with both linear classification and capacity, with the latter scaling exponentially with reservoir size.  

\end{abstract}

\section{Introduction} \label{sec:intro}


Classical models in computational neuroscience are based on rate coding.  The existing state of the art in machine learning (ranging from perceptrons to CNNs and RNNs) also implicitly models rate coding, with the real value at the input or output of a neuron serving as an abstraction of a firing rate.  While there is significant experimental evidence and considerable speculation on the importance of spike timing, computational models are far less established. In this paper, we propose and investigate an idealized model in order to derive insight into the neural codes that can be constructed from spike timing information.

Sensory information (e.g., in visual and auditory stimuli) are inherently embedded into spatiotemporal patterns. 
Experimental evidence in \cite{Cat_visual_cortex,odor_seq} suggests that neurons show dynamical behavior even when the stimulus is fixed: \cite{Cat_visual_cortex} shows that the cat primary visual cortex acts like a fading memory whose current activity contain as much information from previous frame's content as the current frame, while \cite{odor_seq} shows that the network responds differently to odor B when it is proceeded by odor A. This is consistent with the dynamical and state dependent behavior provided by recurrent connections. Thus, recurrent connections, which are the key aspect of the model considered here,
are strongly neuro-plausible candidates for cortical modeling, since they bring about temporal context by providing state-dependent dynamics.  

\begin{figure}[h]
\centering
\includegraphics[width=0.5\columnwidth]{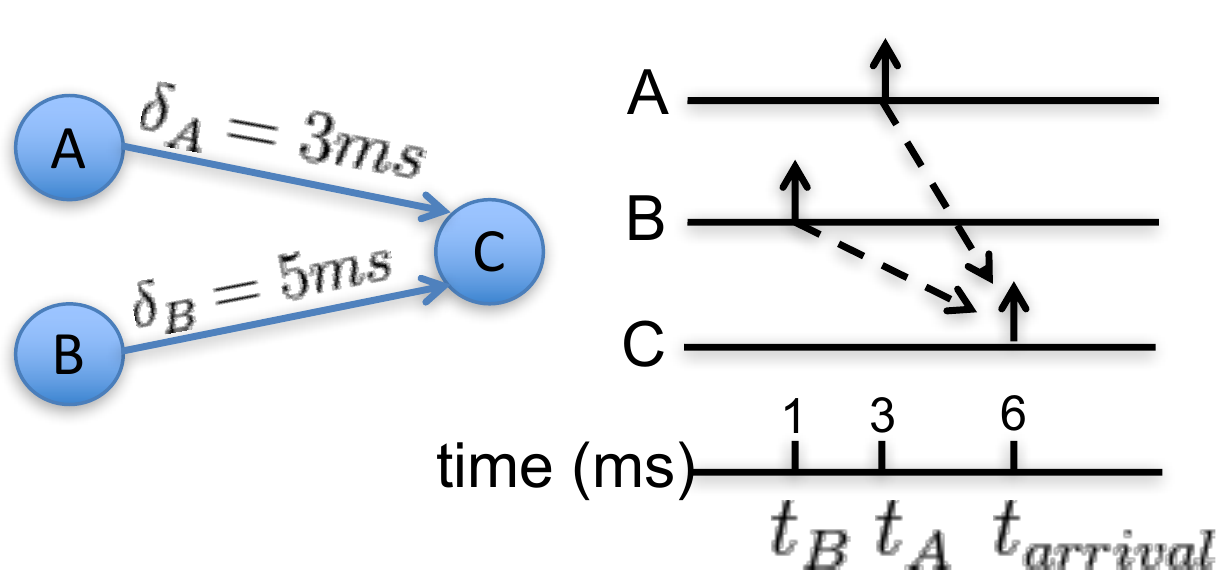}
\caption{Polychronization} 
\label{fig:polychron_example}
\end{figure}

Our starting point is a key observation by Izhikevich more than a decade ago \cite{Izhikevich2006_polychronization}.  He observed that experiments show a great deal of variability in axonal delays, which results in an interesting phenomenon that he terms {\it polychronization.} Figure \ref{fig:polychron_example} illustrates the effect of axonal delays:  Neurons A and B provide inputs to Neuron C through axons with delays 3 ms and 5 ms, respectively.  Suppose that Neuron C acts as a coincidence detector, firing when spikes from Neurons A and B arrive at the same time (within the same 1 ms bin, in our discrete time model).  Given the difference in axonal delays, this happens when the spike from Neuron A is launched 2 ms after the spike from Neuron B.  When we have a network of neurons connected via variable delay axons, interesting patterns of firings in space (i.e., across neurons) and time, which Izhikevich terms {\it polychronous groups,} result from individual spikes initiated at a small number of neurons.

\begin{figure}[h]
\begin{minipage}[ht]{0.48\linewidth}
\centering
\subcaptionbox{}
    {\includegraphics[valign=T, width=1\textwidth]{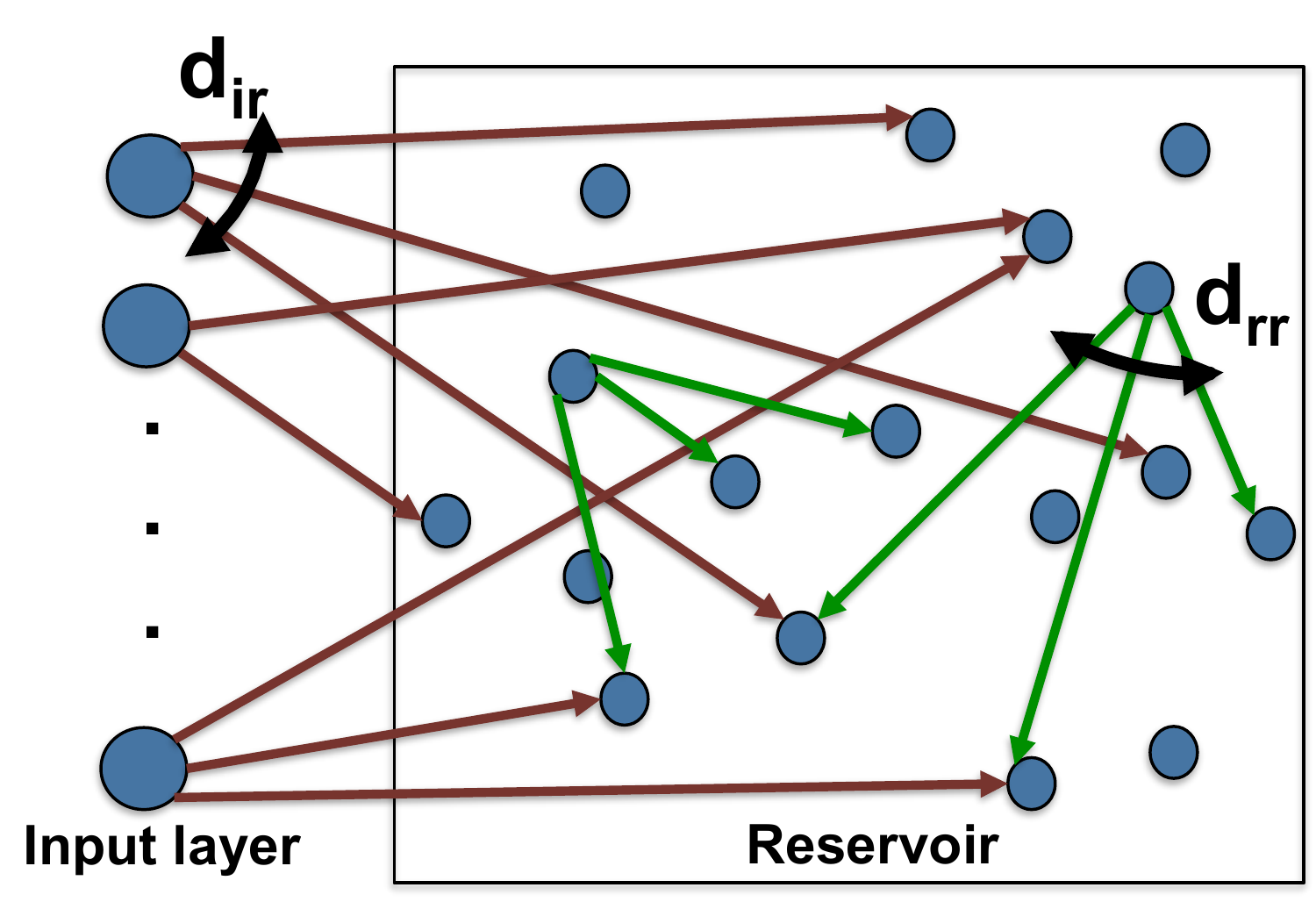}} \hfill
    \end{minipage}
\hspace{0.1cm}
\begin{minipage}[ht]{0.48\linewidth} 
    \centering
\subcaptionbox{}
    {\includegraphics[valign=T, width=1\textwidth]{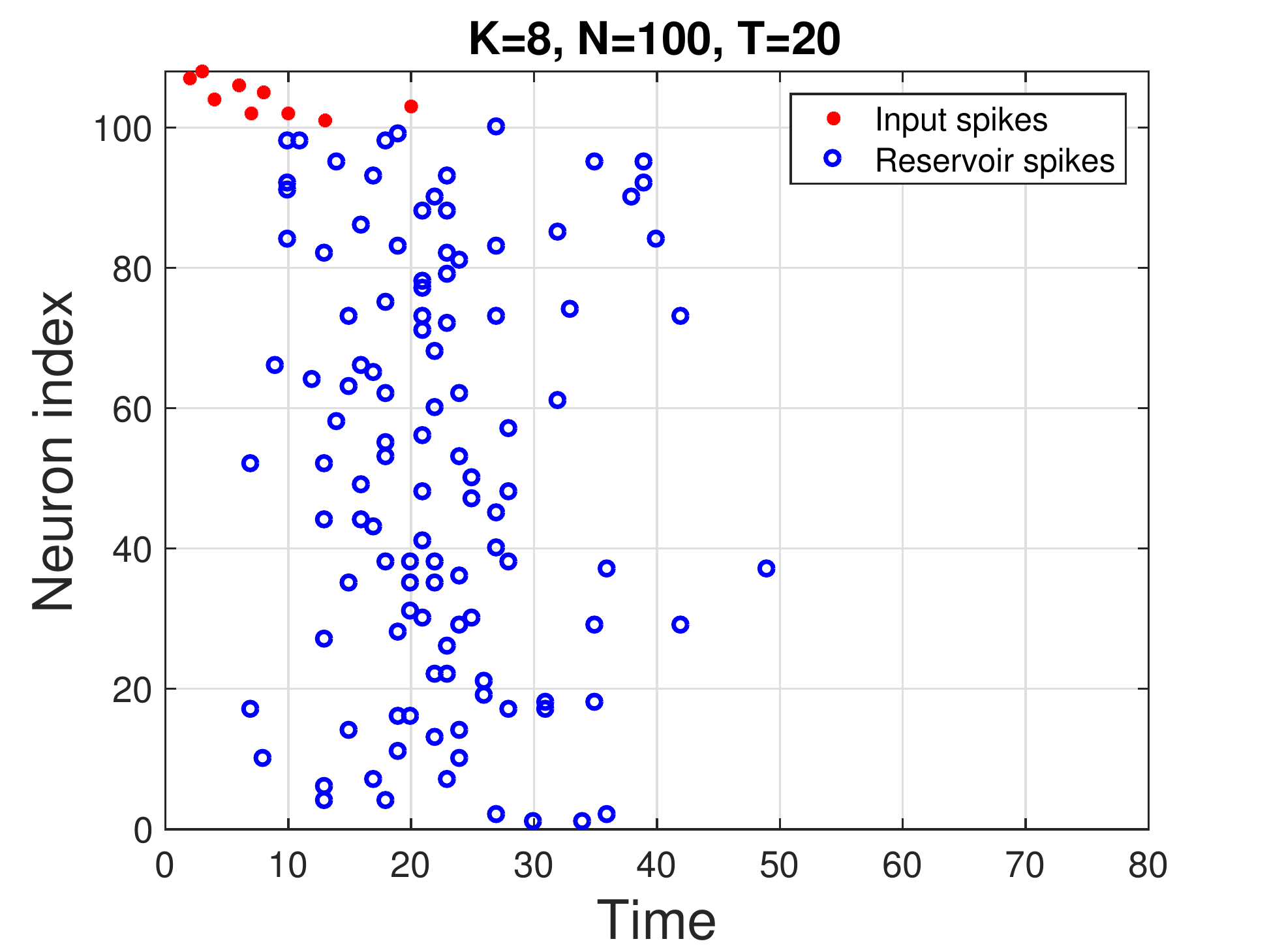}} \hfill  
\end{minipage}  
\caption{(a) System model, (b) Spatiotemporal code}    
\label{fig:Model}
\end{figure}

In this paper, we ask whether polychronization can provide the basis for a concrete computational framework.  In the tradition of information theory, we consider the simplest possible model that captures the basic features of the problem.  We consider a set of $K$ input neurons connected to a reservoir of $N$ neurons, where $N \gg K$,
as depicted in Figure \ref{fig:Model}(a) (the model is discussed in more detail in Section \ref{sec:system_model}). The axonal delays from input neurons to reservoir neurons, and those between reservoir neurons, are randomly chosen from one of $T$ consecutive values. Each reservoir neuron is a coincidence detector, firing when at least m spikes from pre-synaptic neurons fall in the same time bin. For each input spike pattern, there is a corresponding space-time pattern in the reservoir, which we may interpret as an encoding of the input pattern. Figure \ref{fig:Model}(b) shows an example reservoir response to a particular input pattern.

Thus, we have encoded an input space-time pattern into an output
space-time pattern in a higher-dimensional space. The dimension increase from
input to output spaces should separate out the patterns from each
other more than at the input, and hence provides a ``channel code.''
The information embedded in the reservoir's space-time pattern can be read out with various degrees of sophistication. In this paper, we
show that even the {\it simplest} approach in which we integrate the number of spikes for each neuron over a relatively large time horizon (effectively converting the rich dynamics within the reservoir into a rate code) provides 
a powerful neural code that scales ``well'' with reservoir size $N$. This readout mechanism maps the space-time input patterns to spatial output patterns by integrating
the outputs of the reservoir neurons over predefined intervals. The resulting
spatial codeword is a real-valued vector of dimension $N$. We discuss the properties of these spatial codes in this paper. We are interested in how these codes scale with reservoir size $N$, and our main results are summarized as follows:\\
\noindent
$\bullet$ Our analysis reveals that to achieve good properties in terms of linear separability and memorization capacity, the output degree of the input neurons $d_{ir}$, should scale linearly with $N\sqrt{T}/K$, while the output degree of the reservoir neurons $d_{rr}$, can remain constant as we scale the reservoir size. \\
\noindent
$\bullet$ We have verified the above scaling laws in terms of the ability of the neural network to encode spatio-temporal patterns to facilitate classification using a linear classifier. Specifically, we have shown that the network placement of codewords enable it to achieve fundamental benchmarks of linear separability established by Cover \cite{cover1965Geometrical}. \\
\noindent
$\bullet$ We have also verified that, under the above scaling law, the weight and distance properties of the spatial codes scale with
$N$ in a manner similar to that of random codes. This enable us to argue that the number of patterns that can be reliably distinguished
 scales exponentially with $N$.

We observe that, for our simple readout model, small
perturbations in the input pattern (e.g., the change in the timing of
a single spike in one of the neurons) lead to large changes in the
output.  Thus, the system is well-matched to one-shot learning and
memorization (i.e., training and testing using the same patterns).

\section{System model and properties} \label{sec:system_model}


We now describe the system model depicted in Figure
\ref{fig:Model}(a) in more detail, and discuss some basic properties and parameter choices.  As a minimal idealization of
spike timing information, we consider $K$ input neurons, each emitting
a single spike at a time chosen from $\{ 1,...,T \}$. Thus, there are
$T^K$ possible input patterns, corresponding to $K \log_2 T$ bits of
information.
The input can therefore be represented as a $K \times T$ binary array
${\bf U}$ with the $(k,t)$th entry $u_{kt} = 1$ if input neuron $k$
spikes at time $t$ and $u_{kt}=0$ otherwise, where $k=1,...,K$,
$t=1,...,T$.

We are interested in encoding each input spatiotemporal pattern into
the responses of $N$ reservoir neurons.  Each input neuron is
connected to $d_{ir}$ reservoir neurons, chosen at random.  The
reservoir neurons are internally connected to each other, with each
reservoir neuron providing input to $d_{rr}$ other reservoir neurons,
chosen at random.  Thus, the average in-degree of a reservoir neuron
is $\frac{K}{N} d_{ir} + d_{rr}$.  For simplicity, all synapses are
taken to be excitatory, but the role of inhibitory synapses in shaping
neural codes in our setting is certainly of great interest for future
work.

A reservoir neuron fires if and only if at least $m$ incoming spikes arrive in a given time slot. 
We set $m=2$ for all of our numerical results.  

We also make a drastic abstraction of spike timing dependent plasticity (STDP) in our model.
Specifically, we assume that, in the learning phase, all synapses are of equal strength, so that $m$ incoming spikes that line up in time
suffice to make a neuron fire; this enables ``instant learning'' of input patterns.  After the learning phase is over, we may set ``unused'' incoming synapses (i.e., those for which the output neuron has never fired in the slot directly after the arrival of a spike on the synapse) to zero. Incorporating more sophisticated STDP mechanisms compatible with continual learning and forgetting into our model 
are a subject of future work.

\subsection{Neural coding}

While the input patterns span times 1 through $T$, the recurrent
connections in the reservoir imply that the response to an input may
last for more than $T$ time units. In principle, due to reverberant
effects, the duration of spikes from reservoir neurons could be
indefinitely long, but in practice, we can capture most of the relevant information within
a finite horizon that we denote as $T_\mrm{hz}$.  In our simulations, we 
employ $T_\mrm{hz} = 4T$.


 The reservoir ``space-time'' response is a $N \times T_\mrm{hz}$ array ${\bf S}$ with
 $(j,t)$th entry $s_{jt} = 1$ if reservoir neuron $j$ spikes at time
 $t$, and $s_{jt} = 0$ otherwise, where $j=1,...,N$,
 $t=1,...,T_\mrm{hz}$.  
In order to translate the reservoir spatiotemporal information to a purely
spatial domain, we count the firings of the reservoir neurons for the
duration of the horizon $T_\mrm{hz}$, which results in an $N$-dimensional
codeword vector 
${\bf X}$ with its $j$th entry given by
$$
x_j = \sum_{t=1}^{T_\mrm{hz}} s_{jt} ~,~~j=1,...,N
$$
From a computational viewpoint, the mapping from input pattern
${\bf U}$ to ${\bf X}$ is a ``hash'' from temporal coding to rate
coding, resulting in a neural code in a familiar vector space which we
can study with conventional techniques. Since this opens the path for
efficient classical computation and learning, most of our discussion
in this paper is devoted to spatial coding.  
From the standpoint of
neuro-plausibility, there is intriguing experimental evidence
regarding the tuning of synaptic integration within cortex
\cite{Garden2008}.

\subsection{Network Parameters}

In order for the neural code to properly utilize the available dimensions as we scale $N$, and to have the desirable characteristics discussed in Section \ref{sec:code_characteristics}, we must ensure that a large fraction of the reservoir neurons are 
stimulated by each input pattern.  Firings in reservoir neurons occur due to direct stimulation from the input, as well as due to the recurrent internal connections, but an excessive number of firings due to internal connections lead, we have found, to less discrimination across the neural codewords.  
We therefore do not scale the internal degrees in the reservoir with $N$, and scale the degrees of the input nodes to ensure sufficient stimulation. 

The firing probability of a reservoir neuron in direct response to the \textit{input} spikes is equivalent to the \textit{coincidence probability} of spike arrival times denoted as $t_\text{arrival}=t_i+{\bf \delta_i}, \quad i \in\{1,2,...,I\}$ (Figure \ref{fig:polychron_example}), where $I$ is the number of input synapses. In our model, input spikes ($t_i$) and axonal delays ($\delta_i$) are independent random variables with distribution $\mathcal{U}[1,T]$. Hence, $t_\text{arrival}$ has a symmetric triangular distribution over the range $[2,2T]$. 
The coincidence probability for uniform random variables is described by the well known \textit{birthday (collision) problem} \cite{BirthdayProblem}: the probability of having at least one coincidence among $n$ realizations of a random varible distributed as $\mathcal{U}[1,d]$. This probability can be approximated by \cite{BirthdayApproximation}
\begin{equation} \label{birthday1}
p(n;d)=1-e^{{-n(n-1)} \over {2d}}
\end{equation}
The distribution of interest to us is triangular rather than uniform, but the coincidence probability for any finite range distribution is bounded below by that of a uniform distribution over the same range \cite{clevenson1991majorization}.  Thus, the coincidence probability of random variable $t_\text{arrival} \sim \Delta[2,2T]$ for spikes coming through $I$ input synapses is lower bounded by 
\begin{equation} \label{birthday2}
p(I;2T-1)=1-e^{{-I(I-1)} \over {2(2T-1)}}
\end{equation}
Using this expression leads to the rule of thumb that the reservoir in-degree should scale as 
$\Omega(\sqrt{T})$ \footnote{We say that $f = \Omega(g)$ if there exist constant $C>0$ such that $C < |f/g| $}, in order to maintain a constant firing probability. Since synapses coming from input layer are uniformly connected to reservoir neurons, $I$ is the sum of $K$ Bernoulli random variables, with mean $E(I)={{K \over N}d_{ir}}$. Hence, $d_{ir}$ should scale as $\Omega({N \over K}\sqrt{T})$ in order to let the input spikes propagate through the reservoir.  

We consider two regimes for scaling of $K$ and $N$.
In the first regime (Section \ref{sec:separability}), we fix $K$ and let $N$ get large, so that $d_{ir}$ scales as $\Omega(N)$.
In the second regime (Section \ref{sec:memorization_capacity}), we increase $K$ and $N$ with $K.N$ fixed, so that $d_{ir}$ scales as $\Omega(1)$.



\subsection{Chaotic Mapping}

\begin{figure}[h]
\begin{minipage}[t]{0.48\linewidth}
\centering
\subcaptionbox{}
    {\includegraphics[width=1\textwidth]{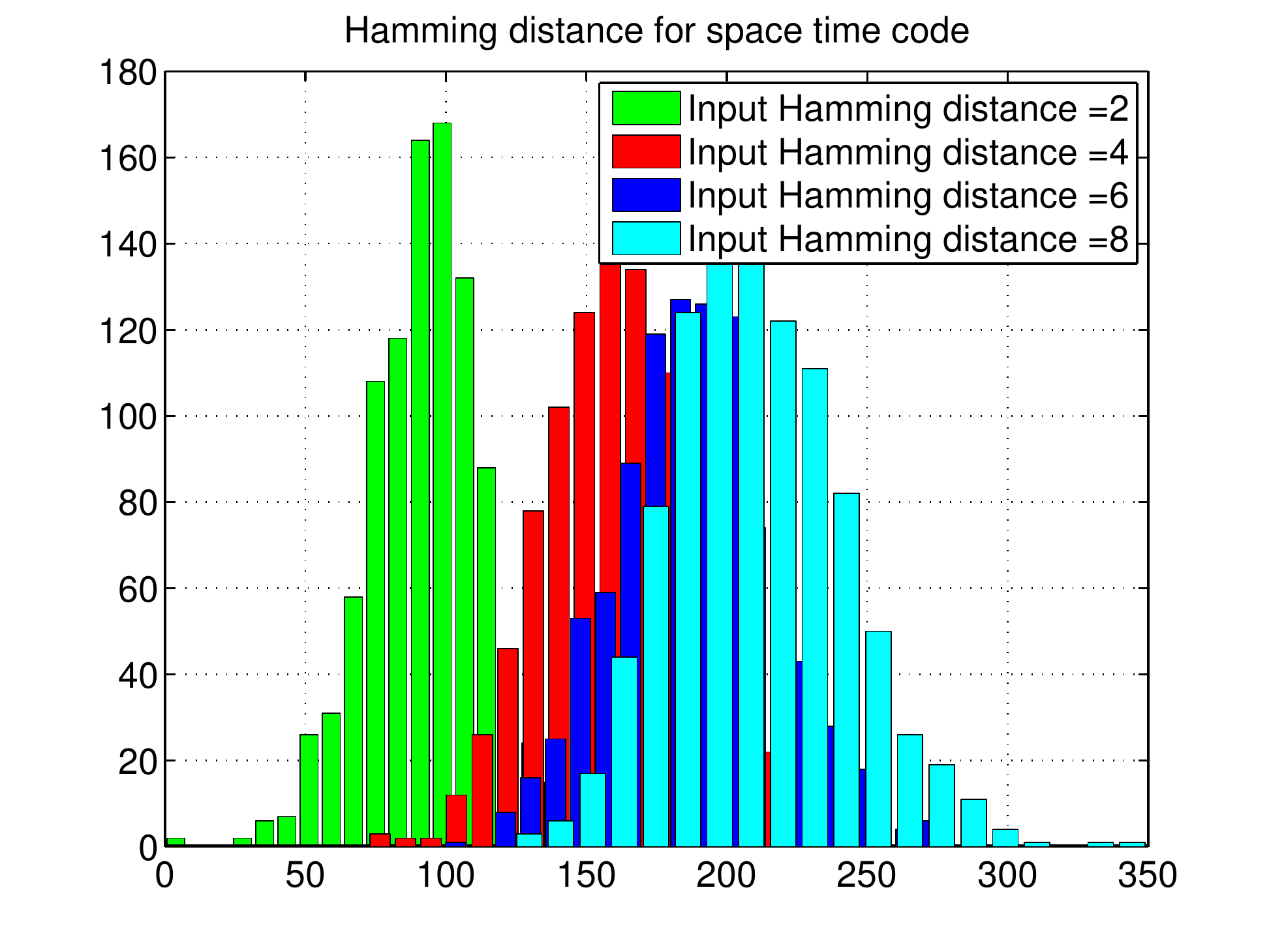}} \hfill
    \end{minipage}
\hspace{0.1cm}
\begin{minipage}[t]{0.48\linewidth} 
    \centering
\subcaptionbox{}
    {\includegraphics[width=1\textwidth]{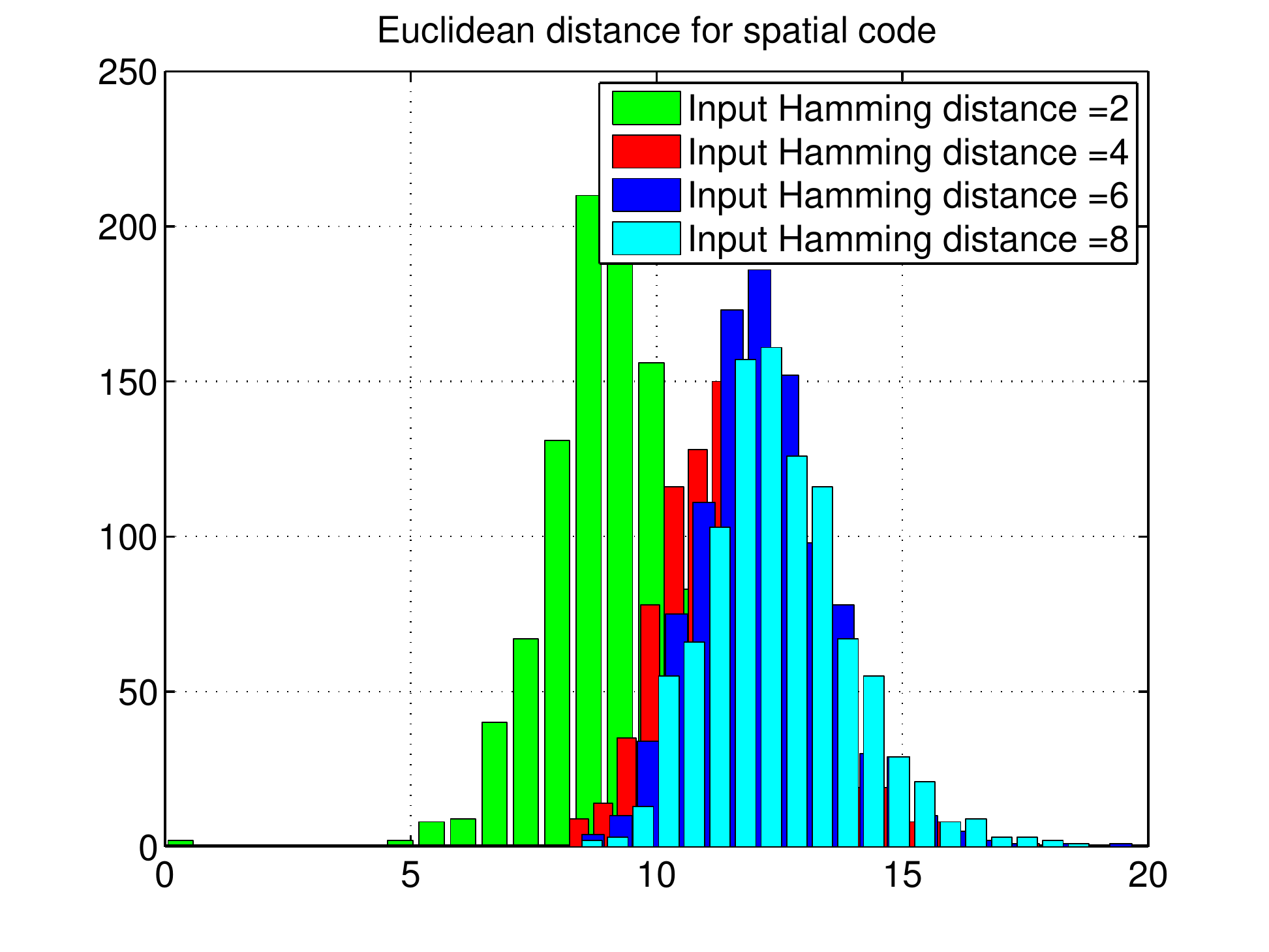}} \hfill  
\end{minipage}  
  \caption{Histogram of (a) Hamming distance of space-time codes (b) Euclidean distance of spatial codes, for different input hamming distances.}
  \label{fig:chaotic}
\end{figure}

Figure \ref{fig:chaotic} indicates that the mapping from input patterns to both the space-time and spatial codes is chaotic, in that the output Hamming distances and Euclidean distances can be large even when the input Hamming distance is small (while the distances for input Hamming distance 2 are smaller than the rest, but there is a substantial fraction of large distances even there).  This sensitivity may be a drawback for generalization, but it is a positive feature for one-shot learning or pattern memorization.
Of course, if we think of our discrete time model as approximating a continuous-time system with a given time resolution, the response will not change abruptly under timing perturbations that are smaller than the resolution.


\section{Code characteristics} \label{sec:code_characteristics}
\subsection{Linear Separability} \label{sec:separability}

From a computational perspective, one can view the reservoir neurons
as mapping the spatio-temporal stimuli applied to the input neurons
into spatial output codewords. In the context of our model, this means
that the neural network is a (nonlinear) map $\phi$ from an input space
of $K$-dimensional vectors (each entry representing a firing time for
the corresponding input neuron) into an output space of
$N$-dimensional vectors (each entry representing the number of firing
for the corresponding reservoir neuron).

The problem of constructing optimal maps from a finite set of input
patterns to an $N$-dimensional output ``feature'' space is addressed
in a classic paper by Cover \cite{cover1965Geometrical}. Cover considers a classification problem
where a set of $P$ patterns represented by features in $\R^d$ are given
binary labels and shows that the fraction $\rho$ of the total set of
$2^P$ possible labels can be classified with a linear classifier is
given by
\begin{align}\label{eq:rho-max}
  \rho\le \rho_{\max}(P,d) \eqdef \frac{\sum_{k=0}^{d}\binom{P-1}{k}}{2^{P-1}},
\end{align}
with equality achieved\ when the $P$ patterns are
in ``general positions in $\R^d$''. By this, we mean that every subset
of $d$ patterns correspond to linearly independent vectors in
$\R^d$. Note that Cover's result in \cite{cover1965Geometrical}
  only considers equality in (\ref{eq:rho-max}), but his proof can easily be adapted
  to show that points not in general positions do not increase the
  fraction of linearly classifiable labels.

\begin{figure*}[!tbp]
  \begin{subfigure}[b]{0.3\textwidth}
    \includegraphics[width=\textwidth]{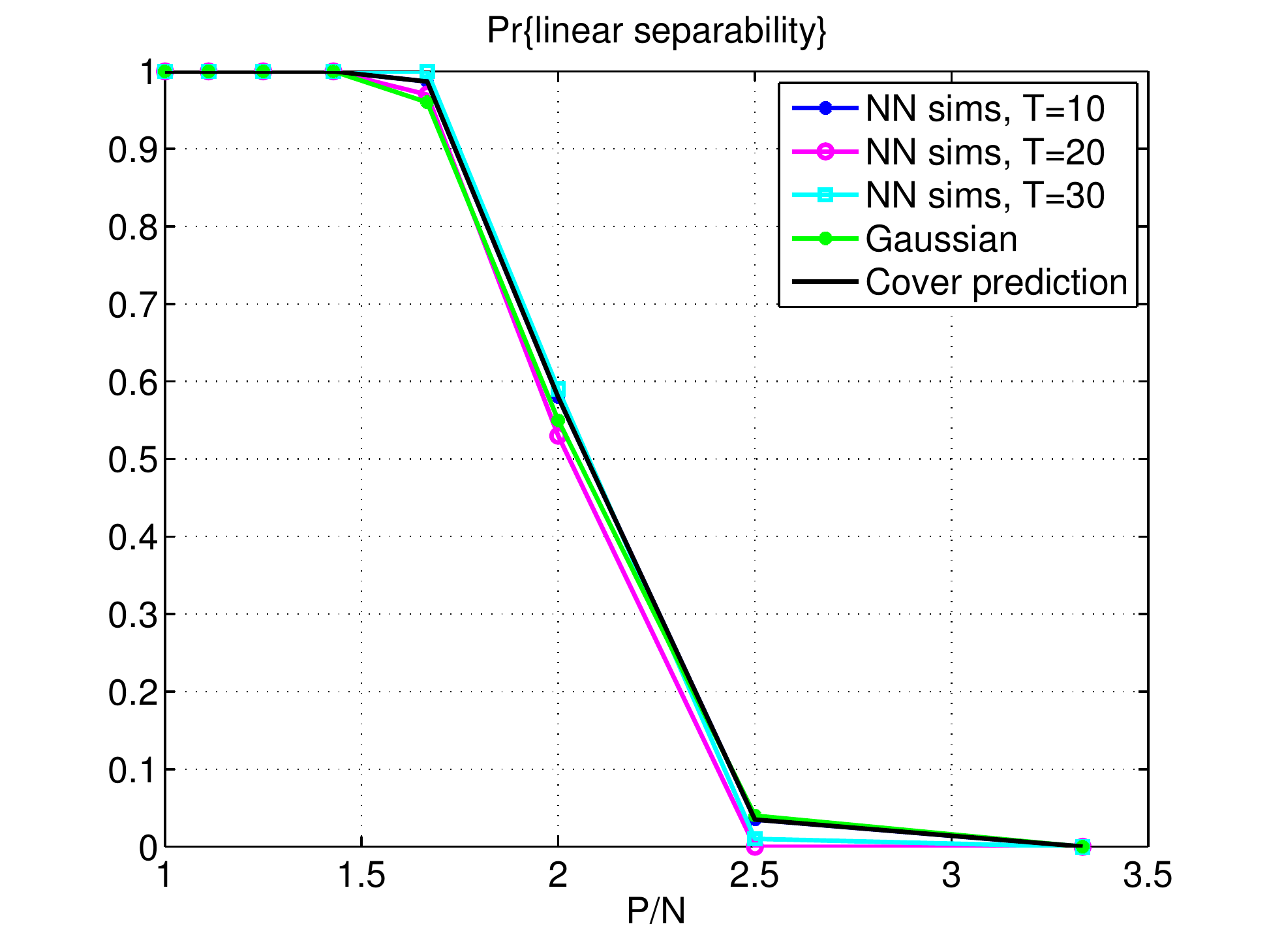}
    \caption{}
    \label{fig:cover_vs_NN_T}
  \end{subfigure}
  \hfill
  \begin{subfigure}[b]{0.3\textwidth}
    \includegraphics[width=\textwidth]{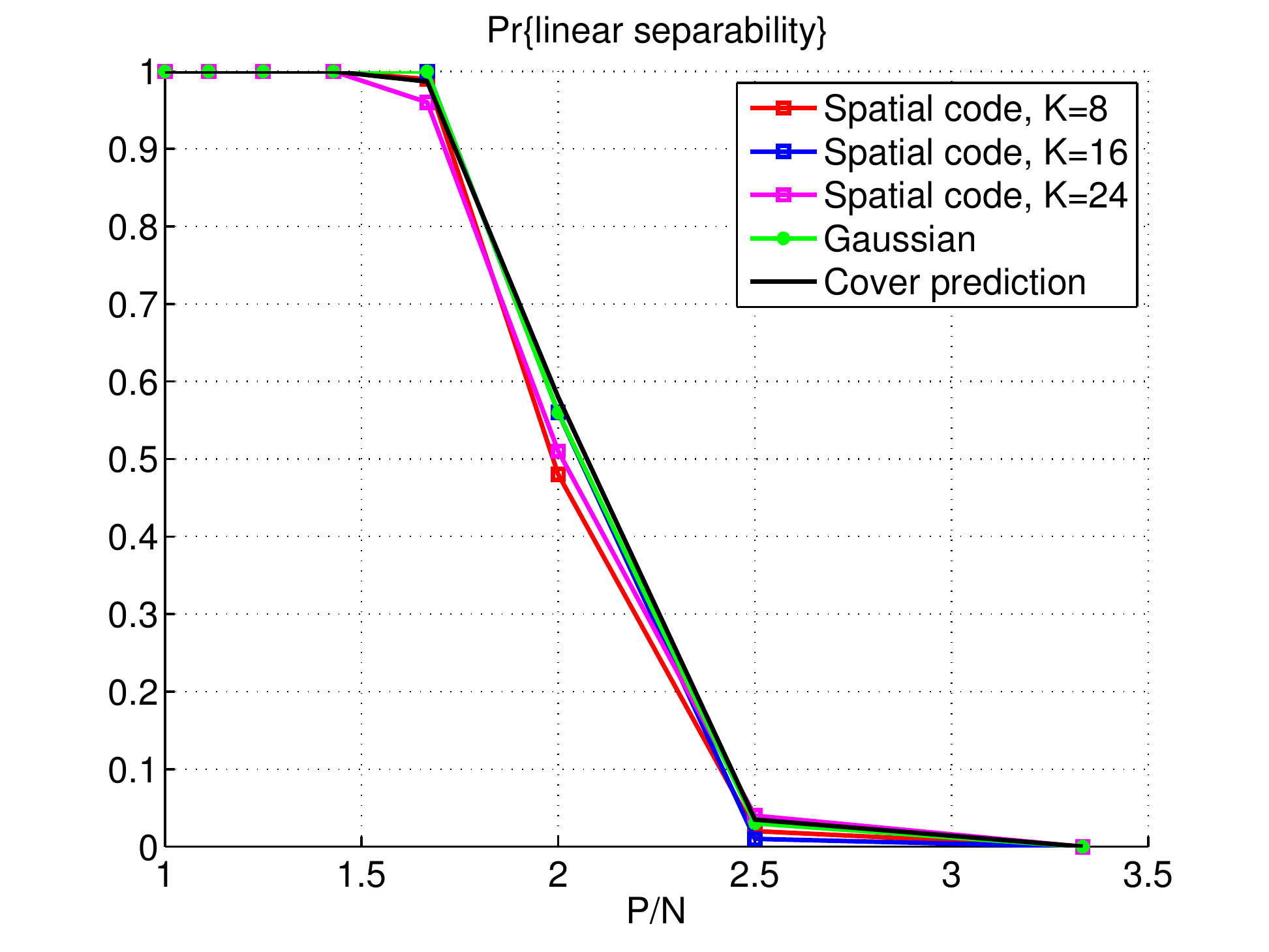}
    \caption{}
    \label{fig:cover_vs_NN_K}
  \end{subfigure}
  \hfill
  \begin{subfigure}[b]{0.3\textwidth}
    \includegraphics[width=\textwidth]{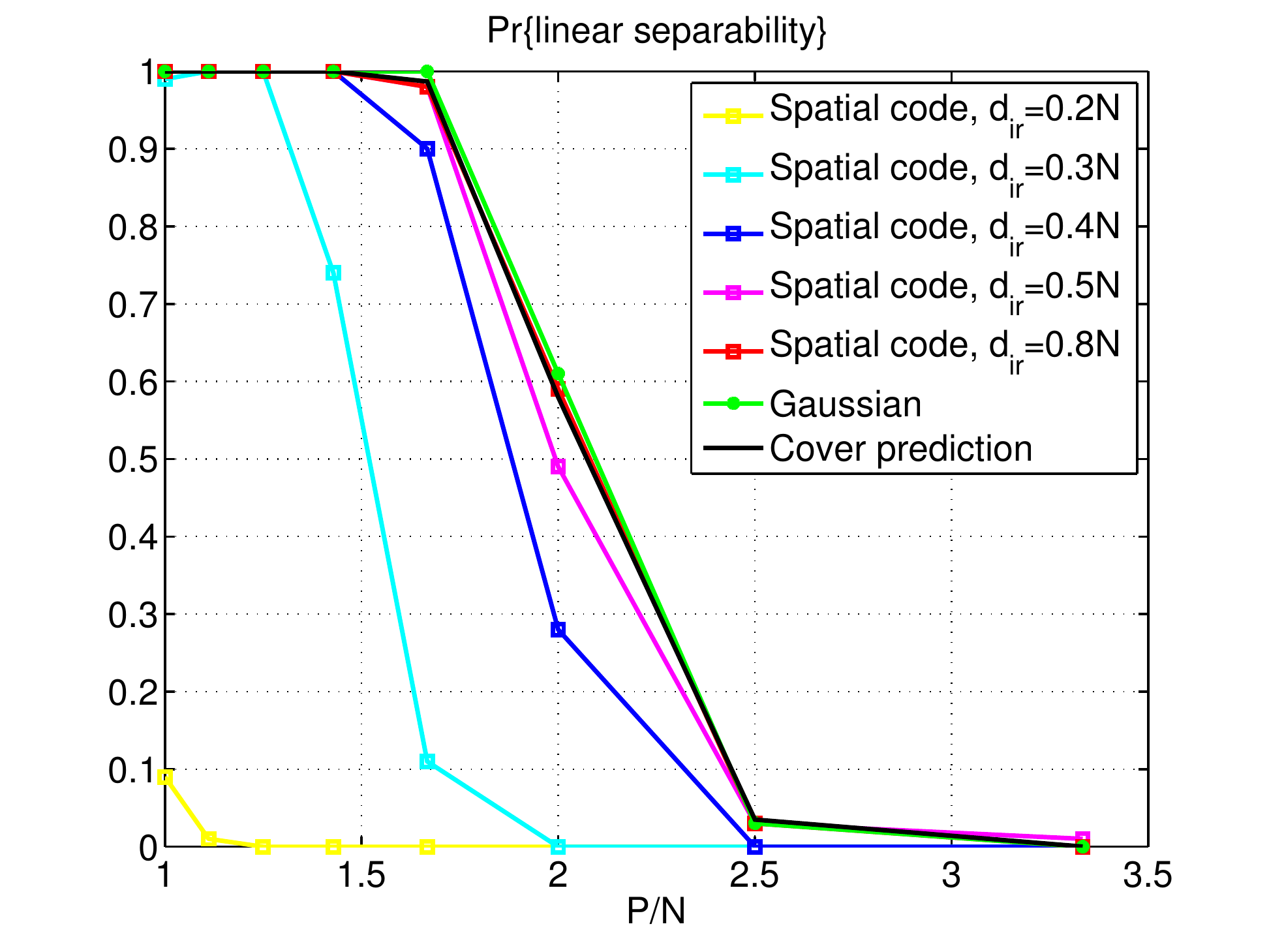}
    \caption{}
    \label{fig:cover_vs_NN_d}
  \end{subfigure}
  \caption{Spatial coding attains optimum separability over a wide range of network parameters. Default network settings are $T=20$, $K=8$, $d_{ir}=0.8N$ and $d_{rr}=4$.}
  \label{fig:cover_vs_NN}
\end{figure*}

An important feature of the function $\rho_{\max}(P,d)$ in
\eqref{eq:rho-max} is that it exhibits a steep change from
values close to 1 for $P<1.5d$ to values close to 0 for $P>2.5d$, as
can be seen in Figure~\ref{fig:cover_vs_NN}, where $\rho_{\max}(P,d)$ is
ploted as a function of $P/d$ (for $d=N$).

Since $\rho_{\max}(P,d)$ is a monotonically increasing function of
$d$, the preceding result shows that one can perform a richer class of
classification tasks by increasing the dimension $d$ of the feature
space. In fact, one can classify essentially any labeling of $P$
patterns represented by features in $\R^d$, provided that
$d>P/1.5$. However, this is only true if we can map our original $P$
patterns into the high dimensional space $\R^d$, \emph{while keeping
the high-dimensional features in general positions.} This excludes, for example,
any linear transformation, which would necessarily introduce
non-general positions when the original patterns lie in $\R^K$ with
$K\ge d$.

Figure~\ref{fig:cover_vs_NN} shows that our neural network
provides an essentially optimal nonlinear mapping of the input spatiotemporal patterns
to spatial output patterns in $\R^N$, in
that it maps the input patterns to general positions in $\R^N$.
This enables linear classification in the output space $\R^N$ for the
largest possible fraction $\rho_{\max}(P,N)$ of possible labels. In order to
obtain this figure, $P=100$ input patterns are randomly generated and
labeled into two classes. These patterns are propagated through
the spiking network, and we then
determine whether or not the patterns can be accurately classified by
a SVM classifier. Figures~\ref{fig:cover_vs_NN_T}
and~\ref{fig:cover_vs_NN_K} compare the fraction of patterns/labels
that can be linearly classified with the upper bound provided by
$\rho_{\max}(P,N)$ (black lines) for several values of the parameters
$T$ and $K$, respectively. For comparison, we also include the
fraction of patterns/labels that can be linearly classified for a set
of random feature vectors in $\R^N$ with iid Gaussian entries (green
lines), which are in general positions in $\R^N$.

Figure~\ref{fig:cover_vs_NN_d} contains a plot similar to that in
Figure~\ref{fig:cover_vs_NN_T}, but instead of presenting curves for
several values of $T$, we fix $T$ and consider several values of the
input-to-reservoir degree $d_{ir}$. This second figure shows that
there is a threshold for $d_{ir}$ below which the mapping is not effective.
Specifically, we see that when each input neuron is connected
to at least $50\%$ of the reservoir neurons, the neural network
essentially achieves optimal separability and can linearly classify
essentially any labeling of $P$ patterns, provided that
$N>P/1.5$. When each input neuron is connected to only $30\%$ of the
reservoir neurons, we see some degradation on performance, but the
network can still classify most labelings, provided that
$N>P/1.25$. However, when the connectivity is down to $20\%$, the
fraction of patterns that can be linearly classified becomes
exceedingly small. 

Thus, while our neural network produces a good code with sufficient connectivity,
there is a threshold below which it is no longer effective.  This is well predicted with the
approximation for reservoir neuron firing probability (due to the input alone) given by (\ref{birthday2}):
the approximations are 0.01, 0.04 and 0.14 for 20\%, 30\% and 50\% connectivity, respectively.
The corresponding numbers via simulations are 0.03, 0.08 and 0.2, respectively, which yield similar qualitative conclusions.
A detailed analysis of the specific threshold
required to trigger ``enough'' activity within the reservoir (which should depend on the reservoir's internal connectivity) is an interesting topic for future work.


\begin{figure*}[!tbp]
  \begin{subfigure}[b]{0.3\textwidth}
    \includegraphics[width=\textwidth]{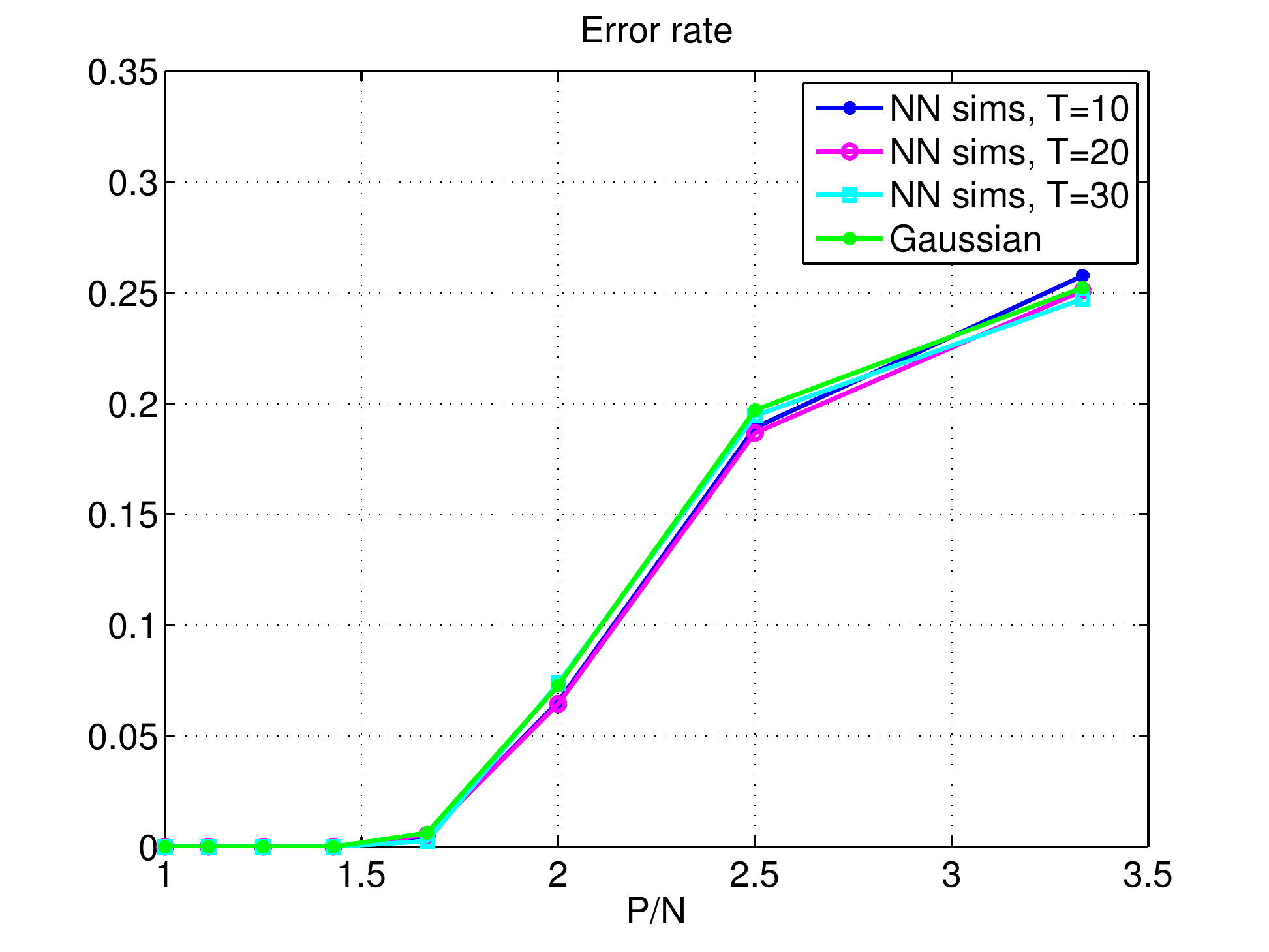}
    \caption{}
    \label{fig:error_rate_T}
  \end{subfigure}
  \hfill
  \begin{subfigure}[b]{0.3\textwidth}
    \includegraphics[width=\textwidth]{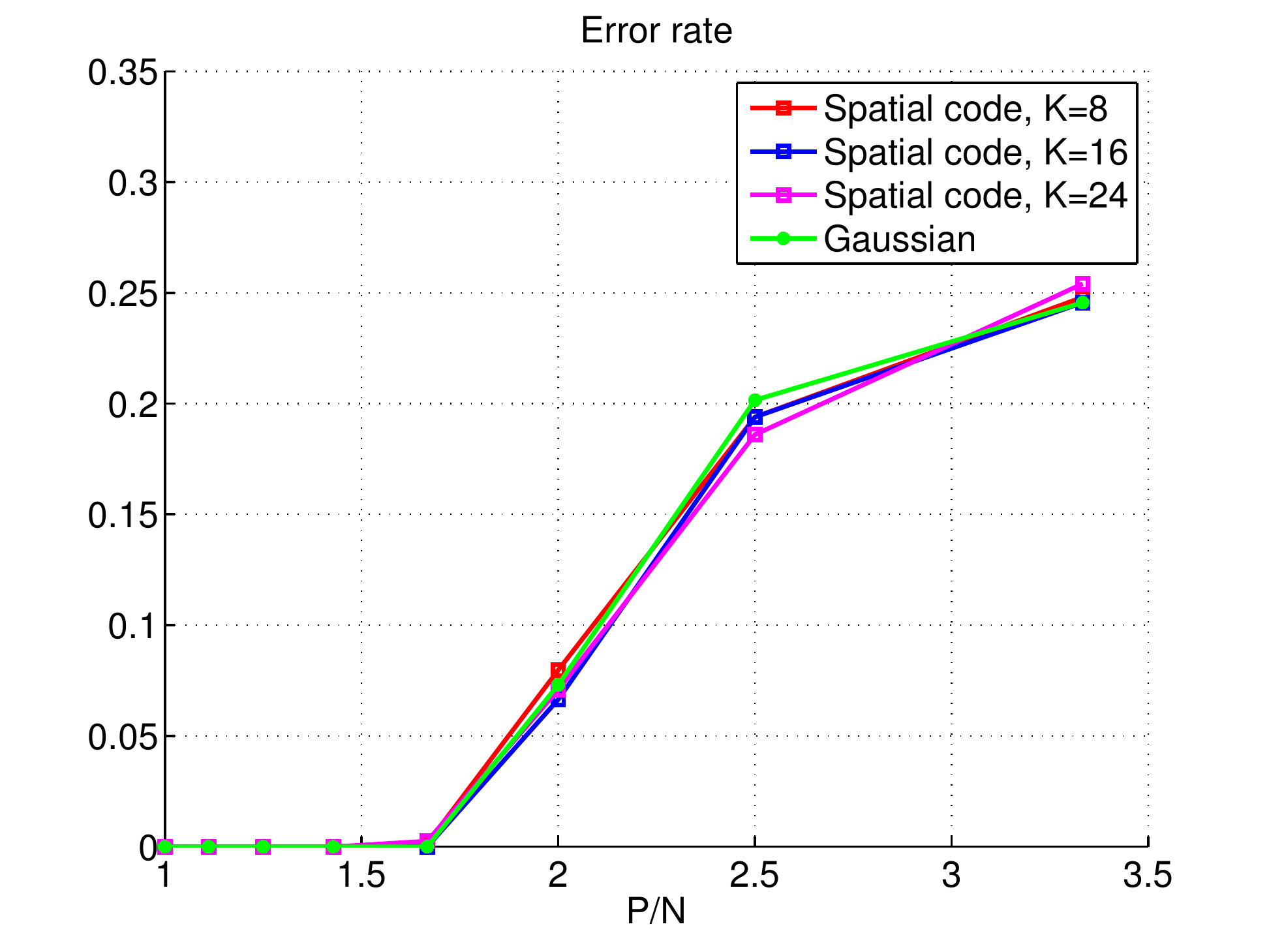}
    \caption{}
    \label{fig:error_rate_K}
  \end{subfigure}
  \hfill
  \begin{subfigure}[b]{0.3\textwidth}
    \includegraphics[width=\textwidth]{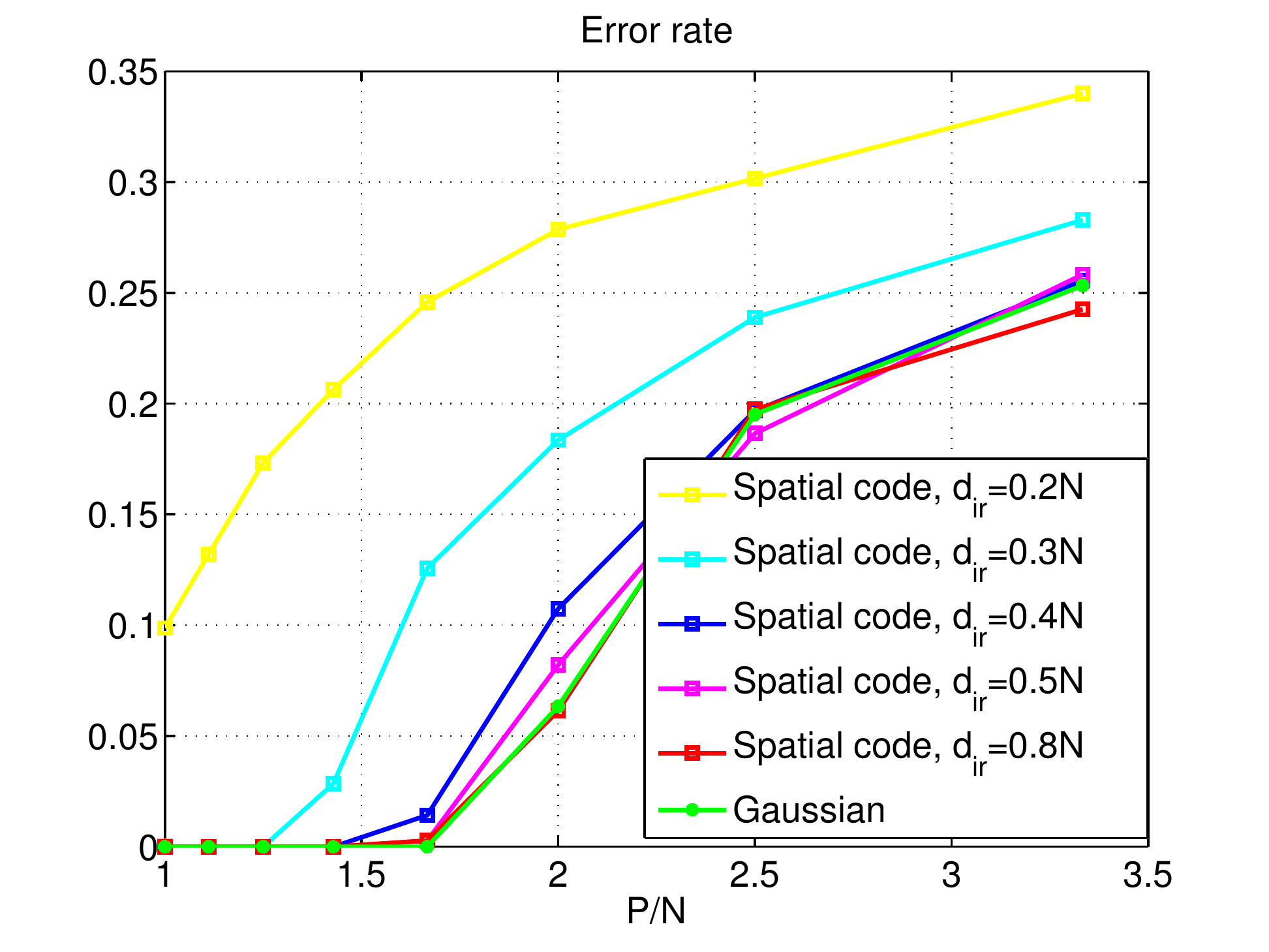}
    \caption{}
    \label{fig:error_rate_d}
  \end{subfigure}
  \caption{Spatial coding's error rate matches with random coding. Default network settings are $T=20$, $K=8$, $d_{ir}=0.8N$ and $d_{rr}=4$.}
  \label{fig:error_rate}
\end{figure*}

The discussion above shows that, under an appropriate
input-to-reservoir connectivity, the neural network is almost optimal
in terms of maximizing the probability of linear
classification. Perhaps not surprisingly, we can show that it also
performs well in terms of achieving a small classification error rate
for the patterns that are not linearly separable in the output space
$\R^N$. To demonstrate this, we present in Figure~\ref{fig:error_rate}
the classification error rates obtained for the same experiments used
to construct Figure~\ref{fig:cover_vs_NN}. We can see that in all the
cases for which we obtained optimality in terms of the probability of
linear classification, the error rates match those that would be
obtained by a set of random feature vectors in $\R^N$ with iid Gaussian
entries.

%



\subsection{Capacity} \label{sec:memorization_capacity}

The linear separability results of the preceding section show that the proposed system produces neural codewords in ``general position,'' similar to those produced by random codes. Of course, the number of linearly separable codewords scales linearly with $N$.  We now remove the constraint of linear
separability, and ask whether the number of patterns we can reliably distinguish scales exponentially with $N$, formulating it as a communication problem as follows:\\
$\bullet$ There are $T^K$ possible input patterns, or messages, corresponding to $K \log_2 T$ bits of information.\\
$\bullet$ Each pattern is encoded into a vector of dimension $N$ using our spatial code, hence our {\it code rate} is
$\alpha = \frac{K \log_2 T}{N}$ bits/dimension.\\
$\bullet$ If we increase code dimension $N$ while keeping code rate $\alpha$ fixed, we have $2^{\alpha N}$ codewords, corresponding to exponential scaling of the memory.  Can we do this while keeping the codewords ``well separated''? 

\begin{figure}[h]
\begin{minipage}[t]{0.45\linewidth}
    \centering
    \includegraphics[width=1\textwidth]{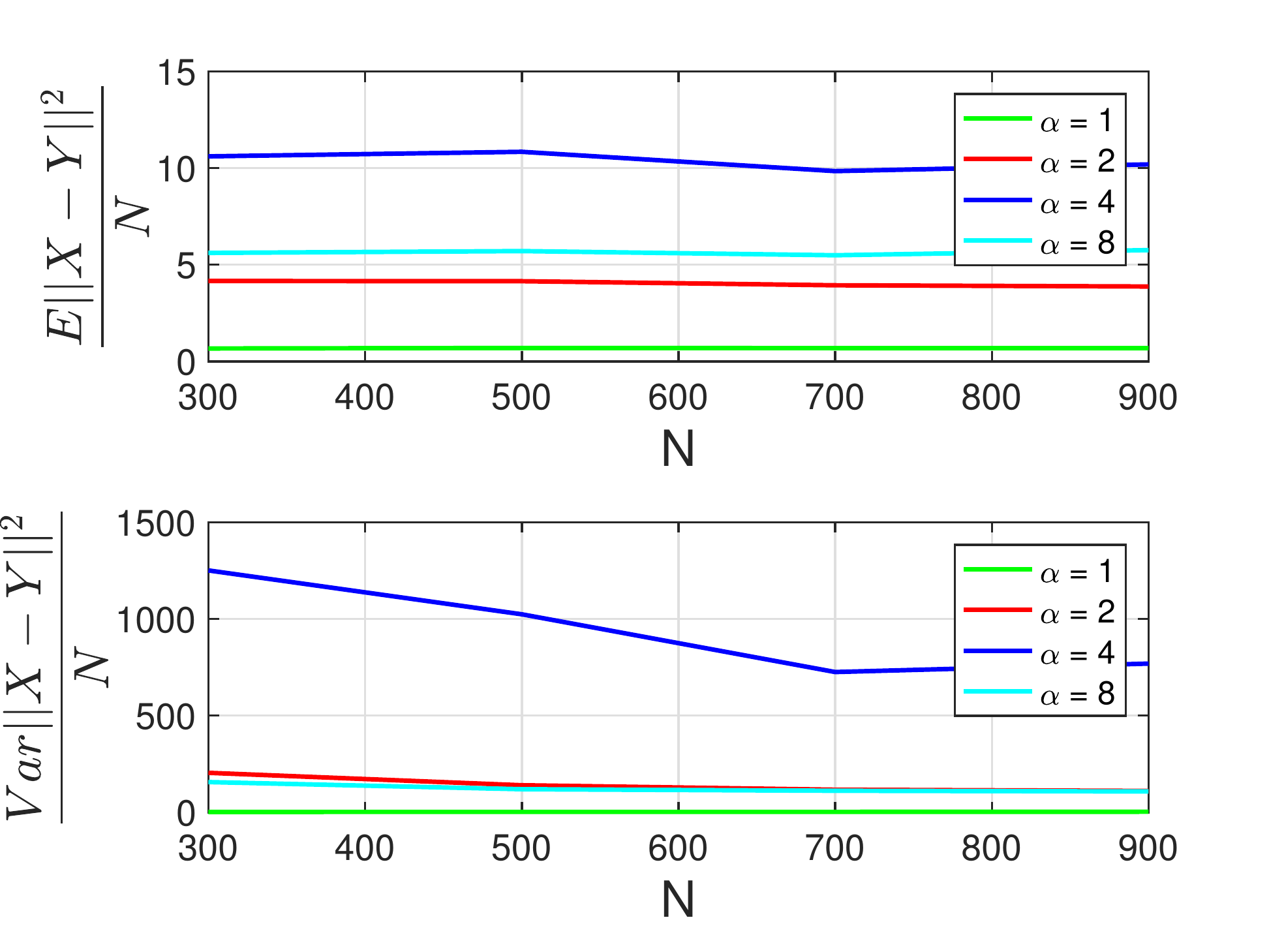}
    \subcaption{}
\end{minipage}
\hspace{0.1cm}
\begin{minipage}[t]{0.45\linewidth} 
    \centering
    \includegraphics[width=1\textwidth]{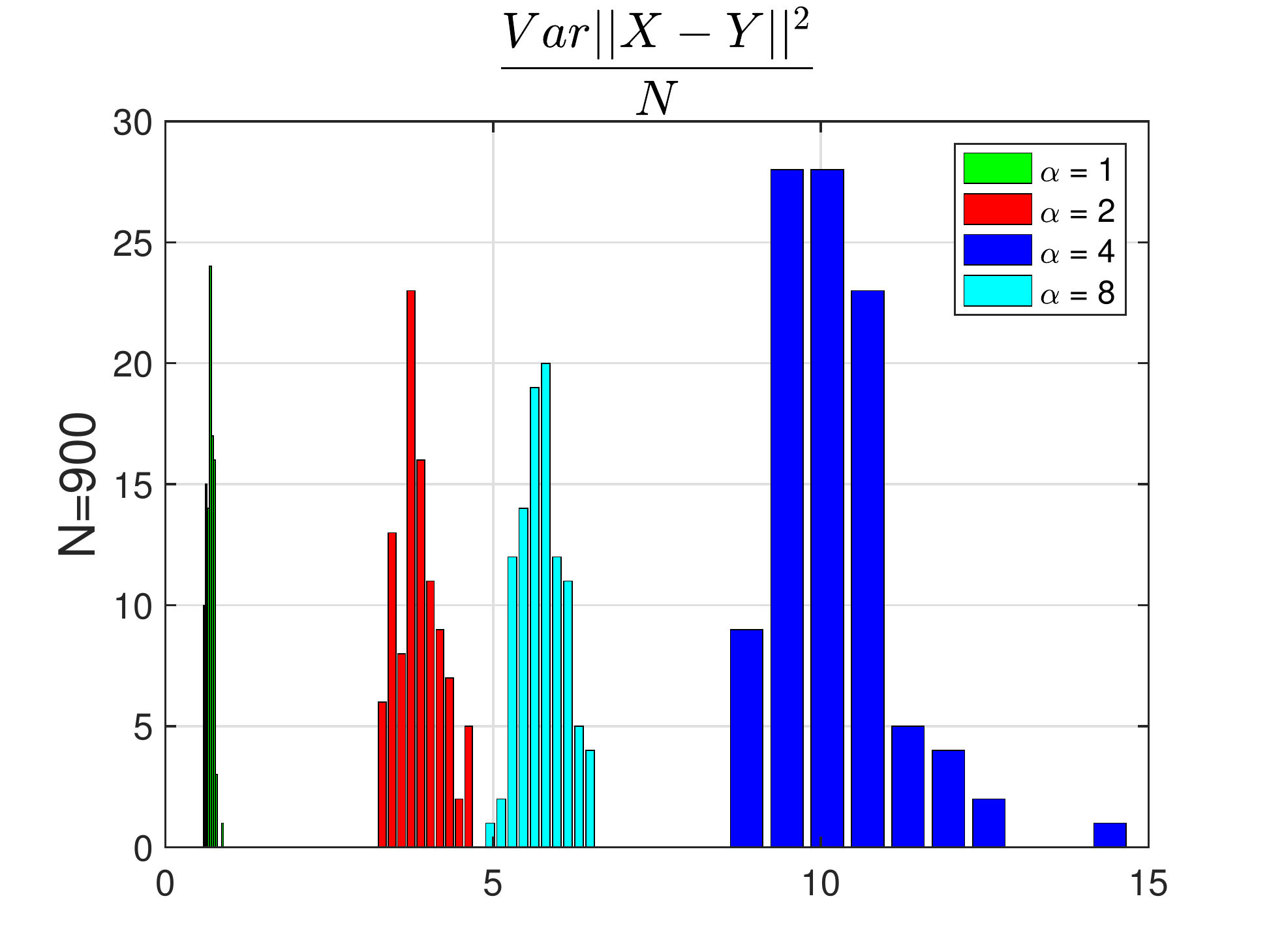}
    \subcaption{}
\end{minipage}     
  \caption{(a) Mean and variance of distance squared scale linearly with $N$, (b) Distance squared concentrates around its mean, and exhibits Gaussian-like deviations from it (consistent with the central limit theorem). Default network settings are $d_{ir}=\sqrt{T}\log{T}$, $T=20$ and $d_{rr}=4$.}   
  \label{fig:distance_properties}
\end{figure}

In information-theoretic terms, ``well separated'' means that, when a codeword is perturbed by noise or other impairments (determined by a channel model), and we use maximum likelihood decoding, then the probability of error (i.e., decoding into the wrong codeword) tends to zero as $N$ gets large. 
For any given input pattern, and corresponding codeword ${\bf X}$, the probability of maximum likelihood decoding to a different codeword ${\bf Y}$ can be bounded by
$e^{- \beta || {\bf X - Y} ||^2}$ for standard noise models such as Gaussian and Poisson.  As we show shortly, the pairwise distance squared between
codewords grows linearly with $N$, concentrating around its mean (Figure \ref{fig:distance_properties}(a)).
Thus, a union bound on the error probability, conditioned on the correct codeword being ${\bf X}$, is given by 
\begin{equation} \label{eq:P_e}
P_e \leq \sum_{{\bf Y} \neq {\bf X}}   e^{- \beta || {\bf X - Y} ||^2} \sim 2^{\alpha N} e^{- \beta \gamma N}
\end{equation}
assuming that $|| {\bf X-Y} ||^2$ concentrates around $\gamma N$.  Thus, the error probability decays to zero if $\alpha \ln 2 < \beta \gamma$.

Figure \ref{fig:distance_properties}(a) gives simulation results showing both the mean and variance of squared distances scale linearly with $N$.
We can now apply Chebyshev's inequality to infer that the squared distances indeed concentrate around their mean, which implies that, for an appropriate
choice of parameters, we can make error probability tend to zero.

Beyond the basic concentration result, it is also worth checking how much our neural codes ``look like'' random codes with i.i.d. components.
Suppose that we have two $N$-dimensional codewords ${\bf X} = (X_1,...,X_N)$ and ${\bf Y} = (Y_1,...,Y_N)$ from a random code.  
If the components are i.i.d. and independent across codewords, the mean and variance of the distance squared will clearly scale linearly with $N$.
In addition, however, $(|| {\bf X - Y} ||^2 - E[|| {\bf X - Y} ||^2])/\sqrt{N {\rm std} \left( X_1 - Y_1 \right)^2}$ tends to a standard Gaussian by the central limit theorem.  Our neural code also appears to exhibit this property: Figure \ref{fig:distance_properties}(b) shows that the histograms of distance-squared exhibit small Gaussian-like deviations around the mean.

We note from figure \ref{fig:distance_properties}(a) that the normalized distance of the codewords, which can be interpreted as robustness of the code against noise, is not a monotonic function of code rate. To explain this behavior, first note that in our simulations, $E(I)=Kd_{ir}/N \, \propto \, \alpha$, which implies that the firing probability of the reservoir neurons is increasing with code rate $\alpha$. For very small $\alpha$ ($\alpha = 0.5$), not enough neurons fire and for very large $\alpha$ ($\alpha = 8$), most of the neurons fire most of the time. Both of these scenarios lead to less discrimination among
codewords. Thus, there exists a sweet spot for code rate (e.g., $\alpha = 4$) that provides the most discrimination by adjusting input-to-reservoir connectivity.

\section{Related work} \label{sec:related_work}
Other prior models trying to utilize spike timing include the tempotron \cite{Gutig2006_tempotron}, which attempts binary classification of different spike timing patterns using a perceptron-like structure, the chronotron \cite{Florian2012_chronotron}, which attempts to train towards a desired output spike timing pattern. Possible mechanisms for realizing such machines are discussed in \cite{Albers2013_associative_learning}.  
There is a rich history and continued research in experimental neuroscience showing the importance of spike timing; see \cite{VanRullen2005_spike_times_make_sense,Rolston2007_precisely_timed,Birznieks2017_spike_timing_matters}, as well as
references in \cite{Izhikevich2006_polychronization,Gutig2006_tempotron,Florian2012_chronotron}. The chaotic nature of the mappings
induced by spike timing has been pointed out in \cite{Banerjee2003}, where these observations are interpreted to cast doubt on whether spike timing provides robust enough signals.
Our model, which is an abstraction of well-accepted spiking neural network models \cite{Gerstner2002_spiking}, embraces the chaotic mappings resulting from spike
timing, and shows that these can produce good ``channel codes'' which could provide the basis for a robust memory. 
Note that Izhikevich also makes an interesting case for the role of polychronous groups in working memory \cite{Izhikevich2010_working_memory}.  Our abstraction is rich enough to produce polychronous groups as in  \cite{Izhikevich2006_polychronization}, without requiring the detailed models for continuous-time neural dynamics used there.

Prior attempts at abstraction include Howard {\it et al} \cite{Howard2010_minimal_model_polychronous}, who consider a network of coincidence detectors similar to ours, but detailed insight into the neural codes generated by such networks is not provided.  In terms of computational models based on polychronous groups, prior work has focused on having at least some synapses with adaptive delays.
Paugham-Moisy {\it et al} \cite{Bengio2008_polychronization} consider supervised learning with a reservoir computing model similar to ours, but with a crucial difference: they allow adjustment of the delays from the reservoir neurons to the readout.  This can, in principle, compensate for the chaotic nature of the map from the input to the reservoir, but adapting delays has drawbacks from the point of view of both computational neuroscience (questionable neuroplausibility) and neuromorphic design (difficulty of implementation).
Izhikevich and Hoppenstead \cite{Izhikevich2009_poly_wavefront} propose a ``polychronous wavefront computation'' model in which transponders fire under a suitable
coincidence between wavefronts initiated by spikes at other transponders, but in order to implement specific function, the location of the transponders (or, in effect, the delays of synapses connecting them) must be programmed.  To the best of our knowledge, ours is the first work to conduct a detailed examination of the neural codes obtained from polychronous groups.


Our model falls within the general framework of reservoir computing, since it has a reservoir with fixed properties, followed by a readout mechanism that has design flexibility.  However, it differs fundamentally from existing reservoir computing
models such as liquid state machines and echo state networks \cite{Maass2002_LSM,Jaeger2001_ESN}.  These are, in essence, fixed nonlinear filters that map the input state to an internal state that is read out after an adaptive linear transformation (adapted to track a target output).  However, the nonlinear mappings in such systems are smooth, unlike the chaotic maps associated with coincidence detection and variable delays in our system.

\section{Conclusions} \label{sec:conclusions}
The new reservoir computing model proposed here enables translation of information from spike timing into ``good'', ''random like'' neural codes in standard vector space.  It is worth exploring architectural variants such as layered reservoirs, or reservoirs with geometric connectivity constraints. The specific readout mechanism we propose is simple, but there could be a number of other ways of going from the binary space-time code defined by the firing patterns of the reservoir neurons to a vector spatial code. The role of inhibitory connections in shaping the code is also an interesting topic. 
While we provide a basic analysis of the neural code, it is of interest to explore its robustness under noise models specifically related to timing:
for example, random Poisson spikes at input or within the reservoir can produce reverberations that are nonlinearly coupled to the input pattern.  

It is interesting to note that spike timing plays a critical role in a number of neuromorphic hardware designs, in the form of the address event representation (AER): analog operations and thresholding encode information in spikes, and the spatiotemporal locations of the spikes are fed to processing and control units via a digital bus.
First proposed in the early 1990s \cite{aer_Sivilotti91,aer_Mahowald92}, AER has been used to build ``silicon retinas'' by a number of research groups \cite{Boahen2000_address_event_representation,aer_nips2005,aer2010}.  Our work raises the question of whether it might be possible to replace explicit digital encoding of the space-time location of spikes, and subsequent digital processing, by more power-efficient analog embeddings and processing. 

\section*{Acknowledgment}

The authors would like to thank Professor Bruno Olshausen for his helpful advice and comments. This work was supported in part by the National Science Foundation under grants CNS-1518812 and ECCS-1608880, the Systems on Nanoscale Information fabriCs (SONIC), one of the six SRC STARnet Centers, sponsored by MARCO and DARPA, and by the Institute for Collaborative Biotechnologies through grant W911NF-09-0001 from the U.S. Army Research Office. \\

\bibliography{ms.bbl}

%
\IEEEpeerreviewmaketitle

\end{document}